# Single-particle entanglement and three forms of ambiguity

*Robert Shaw (shaw2812@gmail.com)*

**Synopsis**

This paper discusses experiments with single-particle systems, some of whose states appear to be entangled. It shows that the interpretation of the experiments in terms of entanglement is ill-defined. Three forms of ambiguity are discussed. The choice of state-space and its dimensions is a matter of taste. There is not an a-priori natural partitioning of the state-space. The observables are not necessarily experimentally accessible and only determined by theory-laden extrapolation from experimental results. These ambiguities need to be addressed in the formulation of any general theory of entanglement.

## I. INTRODUCTION

Entanglement between particles is an idea originally conceived in 1935 in discussions involving Einstein, Bohr and Schrödinger (who coined the term). Sixty years passed before a definition emerged, when Shimony wrote[1]: "A quantum state of a many-particle system may be 'entangled' in the sense of not being a product of single-particle states."

The many-particle restriction was subsequently relaxed, for example in a 2005 paper, "Single-particle entanglement" by van Enk[2] and contested in 2006 by Drezet[3] in "Comment on single-particle entanglement" and physicists have even written about entanglement with the vacuum[4][5]. A more general definition was published by Thaller,[6] "A state of a compound system is called entangled if it cannot be written as a single tensor product of subsystem states. A state in the product form is called unentangled or separable."

As the idea of entanglement has become broader, it has been put forward as the central idea of quantum mechanics: "Entanglement is iron to the classical world's bronze age", [7] say Nielsen and Chuang. In a search for a theory of everything, that transcends any specific realisation, entanglement is being put forward as more fundamental than space or time [8][9][10][11][12][13].

Given its stated significance, it is troubling that there is no consensus about how to define entanglement in a general way that transcends any specific realisation. Bokulich and Jaeger write [14], "The search for a fully general definition of entanglement remains an active area of research."

This note looks at single-particle entanglement to explore some of the issues in pinning down the elusive idea and, in so doing, raises some issues about some of the underpinning notions of quantum theory.

## II. ILLUSTRATIVE EXPERIMENTS

### A. Alice, Bob and Two Photons

Consider an optical Einstein-Podolsky-Rosen experiment. Alice and Bob are located far apart. Alice is equipped with a polarising beam splitter and two single-photon detectors. The beam splitter has the property that it transmits vertically polarised light to D1 but diverts horizontally polarised light to D2. Alice's arrangement is conceptually similar to Stern-Gerlach apparatus in which a magnet serves the role of Alice's beam splitter. Bob has a similar arrangement with detectors D3 (vertical) and D4 (horizontal).

A source located in-between Alice and Bob emits a correlated photon pair. It has the property that the polarisation of either photon is random and the pairs of photons are perfectly correlated. The source sends one photon to Alice and one to Bob. Whenever D1 fires, so does D3. Whenever D2 fires then so does D4. The two situations occur with 50/50 frequency.

Alice's photon system has quantum states $|V_A>$ corresponding to D1 and $|H_A>$ for D2; for Bob the states are $|V_B>$ and $|H_B>$. The composite system of Alice/Bob states can be represented by a tensor product of the Alice/Bob spaces. The correlated state produced by the source has the form:

$$(|V_A, V_B> + |H_A, H_B>) \tag{1}$$

This is just a Bell state and is entangled. Throughout this paper, unimportant normalisations will be neglected.

B. Single-photon mode-entanglement

Suppose a new source sends just one vertically polarised photon to Alice. Detector D1 fires and the quantum state is represented by the vector $|V_A>$. It would appear to be unentangled. Consider as an alternative a Fock representation of the photon $|1> \otimes |0>$, where the numbers inside the kets indicate the photon numbers in the vertical and horizontal modes. This still appears to be unentangled.

Suppose Alice rotates the polarising beam splitter through 45-degrees. Now the detectors D1 and D2 fire equally often. Even though this is a local transformation of Alice's observations, the state now is:

$$(|1> \otimes |0> + |0> \otimes |1>) \tag{2}$$

which would appear now to be entangled in the sense discussed by van Enk and others[15]. Consider now a source that emits a single vertically polarised photon, which then passes through a beam splitter that spatially separates it: half that goes to Alice and half goes to Bob. In this case, either D1 fires or D3, but never both, and D2 and D4 never fire. The composite system of Alice/Bob states can be represented as:

$$|V> \otimes (|A> + |B>) \tag{3}$$

which would appear to be separable into a polarisation substate and a position substate. Now consider the state in terms of mode occupation numbers, for the modes VA, VB, HA, and HB.

$$(|1> \otimes |0> \otimes |0> \otimes |0> + |0> \otimes |1> \otimes |0> \otimes |0>) \tag{4}$$

Under this representation, this state appears to be entangled.

C. Single particle, several varieties

This section considers a thought experiment involving the states of a one-particle system. It is designed to illustrate the concept of entanglement for such systems. All names in this example are fictional.

Families of particles are a conspicuous feature in particle physics; each particle in the family corresponds to a state in *property space*, a term used by Wilczek[16]. Consider two research facilities, Fablabs and Supratron, that use different techniques to create and study new particles. Each facility discovers four particles with identical masses, which they name F1, F2, F3 and F4 (Fablabs), and S1, S2, S3, and S4 (Supratron).

The theoretical physicist Feynberg characterises the Fablab particles with properties Alphaspin and Betacharge, (with linear operators **A** and **B** with eigenvalues ±1). The family of Supratron particles are characterised by Schwingman as having properties Tauspin and Chicharge (with operators **T** and **X** also having eigenvalues ±1).

Figure 1: Particle multiplets in property space, as described by Fablabs and Supratron

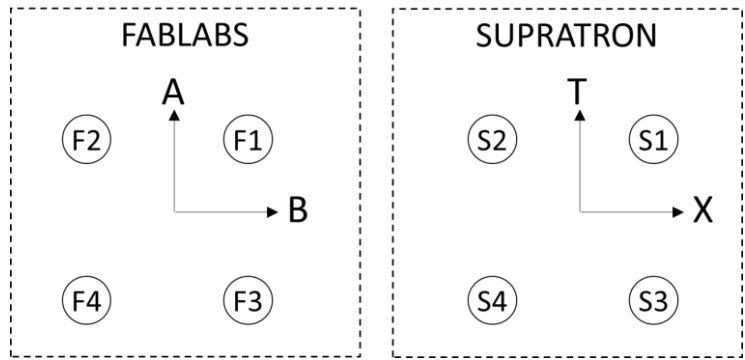

Beams of F particles are studied at Supratron. The F1 beams consist of a 50% mix of S1 and S4. The same is found for F4 beams. F2 beams consist of a 50% mix of S2 and S3; the same is found for F3 beams. Beams of S particles are studied at Fablabs, and a similar pattern is observed.

Feynberg says the S states are entangled and Schwingman says the F states are entangled. Feynberg explains the results by modelling the S states as an entangled superposition of the F states:

$$|S3\rangle \equiv |F2\rangle - |F3\rangle \neq |\rangle_A \otimes |\rangle_B \quad (5)$$

He supports the choice of the term *entangled* by citing examples such as the particles emerging from Stern-Gerlach magnets which are described as states where the spin and position are entangled[17]. Schwingman explains the results by modelling the F as an entangled superpositions of S states:

$$|F3\rangle \equiv |S2\rangle - |S3\rangle \neq |\rangle_T \otimes |\rangle_X \quad (6)$$

The theoretical physicist Dysan analyses both models and concludes that Feynberg and Schwingman explanations are equivalent. The F basis states can be transformed into the S basis states by a unitary transformation:

|    | F1    | F2    | F3     | F4     |
|----|-------|-------|--------|--------|
| S1 | 1/√2  |       |        | 1/√2   |
| S2 |       | 1/√2  | 1/√2   |        |
| S3 |       | 1/√2  | -1/√2  |        |
| S4 | 1/√2  |       |        | -1/√2  |

He concludes that the operators for Alphaspin, Betacharge, Tauspin and Chicharge have matrix representations on the F basis as follows:

$$\text{Rep}_F(A) = \begin{vmatrix} 1 & & & \\ & 1 & & \\ & & -1 & \\ & & & -1 \end{vmatrix} \quad \text{Rep}_F(B) = \begin{vmatrix} 1 & & & \\ & -1 & & \\ & & 1 & \\ & & & -1 \end{vmatrix}$$

$$\text{Rep}_F(T) = \begin{vmatrix} & & 1 & \\ & 1 & & \\ & & & \\ & & & \end{vmatrix} \quad \text{Rep}_F(X) = \begin{vmatrix} & & & 1 \\ & & -1 & \\ & & & \\ & & & \end{vmatrix}$$

$$\begin{vmatrix} & 1 \\ 1 & \end{vmatrix} \qquad \begin{vmatrix} & -1 \\ 1 & \end{vmatrix}$$

The matrix representations for the operators on the S basis are therefore:

$$\text{Rep}_S(A) = \begin{vmatrix} & & & 1 \\ & & 1 & \\ & 1 & & \\ 1 & & & \end{vmatrix} \qquad \text{Rep}_S(B) = \begin{vmatrix} & & & 1 \\ & & -1 & \\ & -1 & & \\ 1 & & & \end{vmatrix}$$

$$\text{Rep}_S(T) = \begin{vmatrix} 1 & & & \\ & 1 & & \\ & & -1 & \\ & & & -1 \end{vmatrix} \qquad \text{Rep}_S(X) = \begin{vmatrix} 1 & & & \\ & -1 & & \\ & & 1 & \\ & & & -1 \end{vmatrix}$$

Dysan determines that both Feynberg and Schwingman are right about entanglement, from their own perspectives. He concludes that no state can be described unconditionally as entangled, and that *whether you view a state as entangled is largely a matter of taste.*

### III. DISCUSSION

Three forms of ambiguity are discussed. These ambiguities need to be addressed in the formulation of any general theory of entanglement.

**The choice of state-space of the system and its dimension has a profound effect on entanglement.** In state-spaces whose dimension cannot be factorised, entanglement is utterly meaningless. This raises the question: what determines the state-space and its dimension? Is it a matter of taste, is it subjective, or is there an objective way of determining the state-space?

Consider the single-photon in section IIB. In a simple matrix representation, it has just two states |V> and |H> and the state-space is self-evidently 2-dimensional, and entanglement has no meaning. However, in a Fock representation the multi-photon state-space is spanned by number-states for vertical and horizontal modes and is infinite-dimensional, so entanglement is possible. The entangled van Enk state in equation (2) exists in a restricted subspace of the full Fock space, which includes the states |0>|0> and |1>|1>, hence the dimension is 4 and the space is separable. Restricting the space further to a single-photon, it is a 2-dimensional subspace of the full space and entanglement has no meaning relative to this reduced state-space.

What determines the state-space? "Quantum mechanics does not tell us for a given system what the state space for that system is" write Nielsen and Huang[18]. It would appear that the state-space and its dimension is a matter of taste. Most physicists are unconcerned by ignoring aspects of a system and representing it in reduced way and moving between different reduced representations. For all practical purposes the choice of state-space is subjective.

**"Partitioning" is a term that has recently been coined to discuss the relationship between a system and its subsystems; it has a profound effect on entanglement.** For example, "To unambiguously define entanglement requires a preferred partition of the overall system into subsystems".[19]

Several authors have suggested that partitioning is "induced" by experimentally accessible observables[20][21][22]. According to this approach the detectors D1, D2, D3 and D4 are what induces the tensor product partitioning in experiment IIA.

This claim is problematic. For the most general sources, the detectors can fire one or more times, depending on the nature of the source, and the space is infinite-dimensional and can be partitioned in many ways.

For a single-particle source, only one out of 4 detectors fires, and the state space is 4-dimensional. This can be factorized into two 2-dimensional subsystems, labelled P/Q and R/S. The detectors provide no assistance in determining how the D-states map to the subsystems. We could guess that the D1,2,3,4-states are: PR, PS, QR, QS; in that case the D1/D2 superposition in equation (2) is separable. Alternatively, we could guess that the D-states are PR, QS, PS, QR and then the state would be entangled. The set of observables is the same in both cases and the claim that experimentally accessible observables "induces" partitioning is not supported.

The essential point is that, even for a given choice of state-space, entanglement is a property of a state relative to a chosen set of subsystems. There is an infinite number of tensor product structures even for a system of two qubits[23], and the choice of a given "partitioning" is a matter of taste. As Zanardi puts it[24] there is "not an a priori god-given partition into elementary subsystems".

**The notion of experimental observation does not resolve the ambiguities of entanglement in the case of "observables" that are not experimentally accessible.** Dirac noted in 1930 that not all measurement operators are experimentally accessible[25]. "Can every observable be measured? The answer theoretically is yes. In practice it may be very awkward, or perhaps beyond the ingenuity of the experimenter, to devise an apparatus which could measure some particular observable, but the theory always allows one to imagine that the measurement can be made."

John Bell commented,[26] "the concept of an observable is a rather woolly concept." He was not just commenting on the philosophical "measurement problem", but the more prosaic matter of what is measured and what is inferred from experimental results.

David Bohm comments,[27] "each observable corresponds to some physical property…in quantum theory the system can either be categorised in terms of a definite position or a definite momentum…in principle there are an infinite number of systems of categorization cutting across both position and momentum. Thus, one can expand the wavefunction in terms of eigenfunctions of the harmonic oscillator or eigenfunctions of the hydrogen atom or in still other ways that will occur to the reader." Observables such as the position of an electron in an atom are not experimentally accessible, and can only be determined by measuring the energy levels of the atom and making theory-laden inferences about the position.

Many properties are inferred not measured, and different theorists may make different inferences as a matter of taste. Neutral Kaons provide an example of the subjectivity of inferred properties. They can be represented on the basis of strangeness or alternatively as Long/Short states. As Griffiths comments,[28] "The neutral Kaon system adds a subtle twist to the old question 'what is a particle?'…the situation is analogous to polarised light. Linear polarisation can be regarded as a superposition of left-circular polarisation and right-circular polarisation…Whether you choose to analyse the process in terms of states of linear polarisation or circular polarisation is largely a matter of taste."

Heisenberg in his essay "The nature of elementary particles" took the position that,[29] "words such as 'divide' and 'consist of' have to a large extent lost their meaning." This note would suggest that the word entanglement has become ambiguous in the case of single particles.

Richard Feynman in the Character of Physical Law[30] said, "Every theoretical physicist who is any good knows six or seven different theoretical representations for exactly the same physics. He knows that they are all equivalent, and that nobody is ever going to be able to decide which one is right at that level, but he keeps them in his head, hoping that they will give him different ideas for guessing."

Entanglement is an exception to Feynman's rule-of-thumb. The different representations of the state-space, the different partitioning, and the different choices of observables, all may lead to similar physical predictions, but when it comes to entanglement, they are not equivalent.

There is something disconcerting about entanglement being "a matter of taste". Many other authors have reached similar conclusions, while expressing their conclusions in a number of ways: "ill-posed problem" [31]; "a convenient notion for expressing more complex ideas" [32]; "entanglement is an inherently relative concept"[33]. For a general theory of entanglement to be formulated, these ambiguities need to be clarified.